# Qualitative assessment of tear dynamics with fluorescein profilometry

*Izabela K. Garaszczuk and D. Robert Iskander*

Department of Biomedical Engineering, Wroclaw University of Science and Technology, Poland

Corresponding author:

Izabela K. Garaszczuk,

Department of Optics, Optometry & Vision Science,

University of Valencia,

Calle del Doctor Moliner, 50 - 46100 Burjassot, Spain,

E-mail: Izabela.garaszczuk@uv.es

**ABSTRACT**

Purpose: To describe a new methodology for tear-film dynamics assessment by observing fluorescein decay rate over time and to understand the relationship between the newly defined tear fluorescein washout rate (TFWR) and other measures of the tear film behaviour.

Methods: Forty subjects (24F/16M) aged (mean ± standard deviation) 31.8 ± 14.2 years volunteered for the study. It consisted of the review of medical history, McMonnies questionnaire (McMQ), slit lamp examination, and TFWR using a newly-developed fluorescein profilometry. The repeatability of TFWR measurements was assessed. TFWR estimates were contrasted against patient age, McMQ score, daytime, fluorescein tear film breakup time (FTBUT), tear meniscus height (TMH) and blink frequency.

Results: Mean repeatability of the method was 28.13 ± 9.59%. The group mean TFWR was 39 ± 23% at 30-s mark after the beginning of measurements, ranging from 1.4% to 83%. This indicates that TFWR is highly subject-dependent. Statistically significant correlations were found between the percentage TFWR and McMQ score ($r^2 = 0.214$, $p = 0.001$) as well as FTBUT ($r^2 = 0.136$, $p = 0.009$). No statistically significant correlations were found between TFWR and age, daytime, TMH, and blink frequency.

Conclusions: Fluorescein profilometry allows clinicians to follow dynamic changes in the tear film on the entire ocular surface and may be used for qualitative assessment of the tear film dynamics.

**INTRODUCTION**

A well-balanced consistency of tear film and effective tear dynamics are essential for ocular surface health and function. The tear dynamics in a healthy eye consists of well-functioning tear production, its distribution and retention on the ocular surface, turnover, elimination through the nasolacrimal system, and evaporation and absorption into surrounding tissues [1]. These processes are regulated by lacrimal functional unit, which is composed of the lacrimal glands, Meibomian glands and conjunctival goblet cells, the ocular surface and interconnecting innervation. Isolated tests of tear dynamics components often poorly correlate with patient symptoms or with other clinical and laboratory tests [2,3].

Tear turnover or tear clearance rates are temporal measures of all dynamic processes occurring in the tear film and they are physiological measures of the integrity of the lacrimal system and tear exchange on the ocular surface [1,4–9]. Tear turnover rate (TTR) is most commonly evaluated with fluorophotometry [5–7,10–13], which provides a direct quantitative assessment of tear flow, drainage and turnover whilst tear clearance rate (TCR) is assessed with tests based on fluorescein clearance tests [8,9,12,14–17]. There are simple alternatives to fluorophotometry that include modifications of the Schirmer test with or without anaesthesia [5,12,16,18] and colorimetric analysis of tear meniscus [15]. TTR and TCR measures take into account several factors affecting tear film and were shown to be the indirect measures of dry eye associated with ocular surface inflammation, ocular irritation, severity of ocular surface disease, Meibomian gland dysfunction or decreased ocular surface sensitivity [4,8,9,14,15,17,19]. Also, factors associated with age, such as conjunctivochalasis, lid laxity, tear flow functional obstruction, and blink abnormalities, may all contribute to delayed tear turnover [9,17,20]. It is also proven to be reduced in patients with symptomatic dry eye [21]. As shown in [5], tear turnover is proportional to the sum of the effects of tear secretion, transudation of fluid through the conjunctiva, tear drainage through nasolacrimal duct, evaporation and conjunctival and corneal permeability.

TTR and TCR measures evaluate tear film dynamics within longer period of time of up to 30 min. On the other hand, a method for observing the dynamic changes in tear meniscus morphology by anterior segment optical coherence tomography (OCT) was recently proposed to study the early-phase of tear clearance as a function of age [22]. In that method, saline solution was instilled to the eye rather than the fluorescein and changes of tear meniscus height over time were observed.

The main goal of this study was to introduce a new, clinically suitable method to assess early-phase fluorescein decay with fluorescein profilometry. Unlike other methods, this approach follows tear dynamics on the whole surface of the eye. The method was inspired by fluorophotometry and differs in terms of both the methodology and characteristics of the phenomena observed. Accordingly, instead of using Tear Turnover Rate, as is typically assessed with the standardized fluorophotometric technique, a new term "Tear Fluorescein Washout Rate (TFWR)" was adopted, as defined in the Methodology section.

## METHODOLOGY

### Fluorescein profilometer working principle

Only one channel of the Eye Surface Profiler (ESP, Eaglet Eye, The Netherlands) was used for TFWR measurements [23]. The instrument is run in a video mode, in which it is possible to permanently switch on the projected grid and record dynamic changes in a series of acquired images. The TFWR is defined as the percentage change of the average image intensity, constrained to the anterior eye surface area, per 30 s.

### Subjects

Forty subjects (24F and 16M) aged (mean ± standard deviation) 31.8 ± 14.2 years (ranged from 21 to 70) volunteered for the study. The study adhered to the tenets of the Declaration of Helsinki. Informed consent was obtained from all subjects. Exclusion criteria included regular contact lens wearers, subjects with signs or symptoms of eye dryness or inflammation, subjects recovering after surgery or with any tear flow impairments and systemic diseases.

The study protocol consisted of the review of medical history, McMonnies questionnaire, slit lamp examination (assessment of eyelids, ocular adnexa and anterior eye surface for sings of irritation or tear flow impairment, lid-parallel conjunctival folds (LIPCOF), fluorescein tear film break-up (TFBUT), meniscus height, blink frequency estimation), and TFWR estimation using the proposed fluorescein profilometry. The temperature in the laboratory was stable and monitored. The mean temperature was 24.6° ± 1.4° degrees Celsius and mean humidity was 26.1 ± 4.4 [%RH]. Following the evidence that TTR and most probably TFWR could vary with daytime [13], the time of subject's awakening was noted and the period between subjects' awakening time and the beginning of the measurement ($\Delta t$) was taken into account.

**Data acquisition**

ESP projects a diffusely emitted in the tear film, when fluorescein is instilled to the eye. In theory, the situation where fluorescein was fully absorbed by the cornea, thus leaving no fluorescent dye in the tear film, would result in no image being observed with the profilometer as the vast part of the observed image, representing conjunctiva, does not reflect light. Based on this observation we assume that the measurements are not biased by the corneal permeability to fluorescein. In this study, ESP was used in an unconventional way to observe dynamic changes occurring in the tear film after fluorescein instillation on the whole eye surface.

For the ESP to deliver accurate and repeatable measurements of the anterior eye surface, two important factors have to be considered:

(1). The instrument has to be situated at the optimal acquisition position with fixation spots and illumination spots aligned;

(2) The anterior eye surface has to be uniformly covered with a mixture of tear film and fluorescein.

In ESP, the optimal acquisition position corresponds to the near (to the operator) end of the instrument's depth of focus (DoF). ESP is equipped with three focus aiding tools to guide the operator to this location. Some practice is necessary to master the acquisition routine and arrive at accurate and repeatable results. The best measurement practice to assess corneo-scleral topography with ESP has been previously described [23].

For the purpose of the study, the ESP's software was equipped with one additional function. Instead of acquiring one set of two projected images, necessary for topography assessment, ESP was enabled to record the video of the eye with observed patterns projected on the eye surface. To not to induce reflex tearing, only one of the two diodes was used to illuminate the eye surface.

The best measurement practice consists of:

•Aligning the instrument with the help of the centration tool;

• Bringing the instrument to the range of DoF with the help of the optical focussing tool;

• Setting the instrument at the near end of the DoF range with the help of the software focusing tool;

• Instilling fluorescein with the ESP blocked in its position;

• In addition, performing small adjustments of focus if necessary, asking the subject to blink naturally and look at the ESP's focusing cross;

• Recording the video until the projected image is barely visible in the tear film, and manually adjusting the position of the instrument to correct for small head and eye movements.

During the ESP measurements care should be taken to instil a sufficient amount of fluorescein onto the eye as this is essential to achieve good contrast of the images. It is also very important for the subject to blink a few times to distribute the dye on the eye surface, avoiding the nonconfluence in the fluorescent pattern that may occur and distort the results. On the other hand, the eye should not be flooded with artificial tears as this will lead to tear pooling and uneven distribution of fluorescent dye in the tear film.

A drop of lubricating ophthalmic solution from a container that allows the application of a specific amount of 0.1% sodium hyaluronate (Hylo-Parin, UrsaPharm, Poland, ursapharm.com) was used to wet the fluorescein sodium 1 mg ophthalmic sterile strips (BioGlo, HUB Pharmaceuticals) which was applied lightly to the subject's lower bulbar conjunctiva. This particular choice of the solution was derived by an earlier study (result not published) in which the effect of solution viscosity was studied that showed 0.1% sodium hyaluronate was good trade-off between the measured corneo-scleral coverage and additional thickening of the tear film. Subjects were asked to blink naturally several times to distribute the dye and then the video was recorded. The approximate acquisition time was up to one minute. The fluorescein was applied in a standardized manner, identical to the one that was incorporated for tear film break-up time estimation. The subject was requested to blink gently three times to distribute the dye before taking the measurement.

Based on ESP working principle described above, the diffused image intensity is proportional to the amount of fluorescein in subject's tear film. The image was visible as long as the fluorescein was present in the subject's tear film. The subject was instructed to focus on the instrument's fixating cross. The recording of the pattern through an in-built yellow filter allowed observation of the fluorescein image intensity decay. The dye was gradually replaced with new tears and patient was allowed to blink freely. The fluorescence intensity decay curve was obtained using a custom written MATLAB (The Mathworks, Natick, MA) programme, which calculated the mean intensity of the image within the defined area, corresponding to the anterior eye surface (as shown in Fig. 1.), for every frame of the captured video. The two main steps of estimating the fluorescence intensity decay curve are:

(1) Generation and application of the mask that extracting the area of analysis (AOA);

(2) Calculation of mean image intensity in the AOA, to which no weighting is applied. An illustrative curve of intensity decay measured on one of the subjects is presented in Fig. 2.

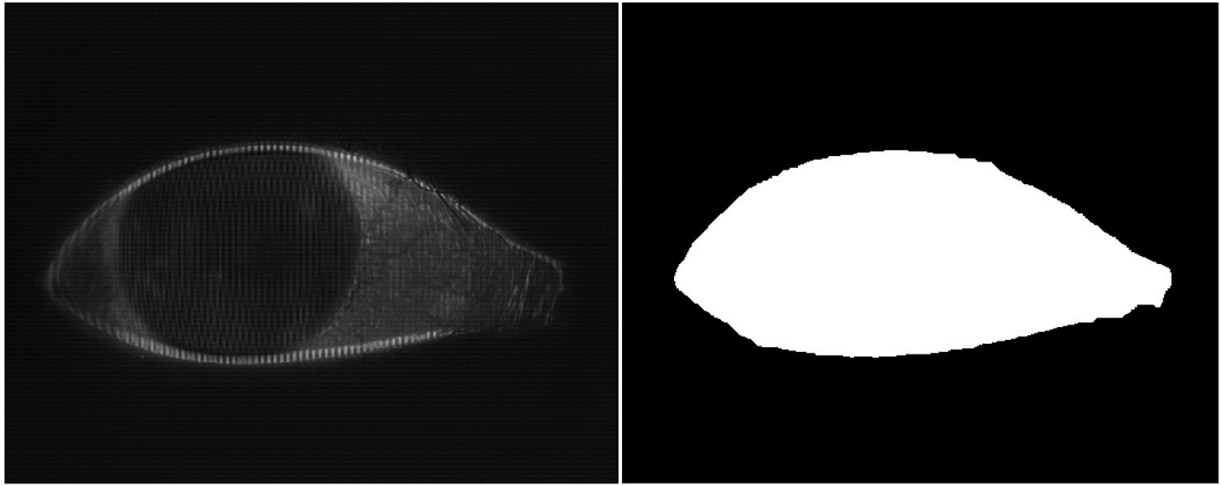

**Fig. 1.** An illustrative frame from the ESP video sequence (left) and the corresponding demarcated area of analysis (right).

The monophasic exponential model of the image intensity decay was assumed, $I(t) = Ae^{-\beta t}$. After removal of the signal artefact due to blinks, the amplitude A and the decay constant $\beta$ were estimated using linear in parameters least-squares procedure by taking first the logarithm of the model, that is, $\log(I(t)) = \log(A) - \beta t = \alpha - \beta t$, where $\alpha = \log(A)$.

A time varying TFWR was defined as:

$$TFWR(t) = \frac{I(t)}{I(0) \times t} 100 \ [\%/min].$$

The TFWR was estimated as a percentage drop of image intensity after t= 30 [s]. The margin was chosen, taking into account that the shortest video from those assessed in the study had a duration of 40 s. The time margin is marked on Figs. 2 and 3.

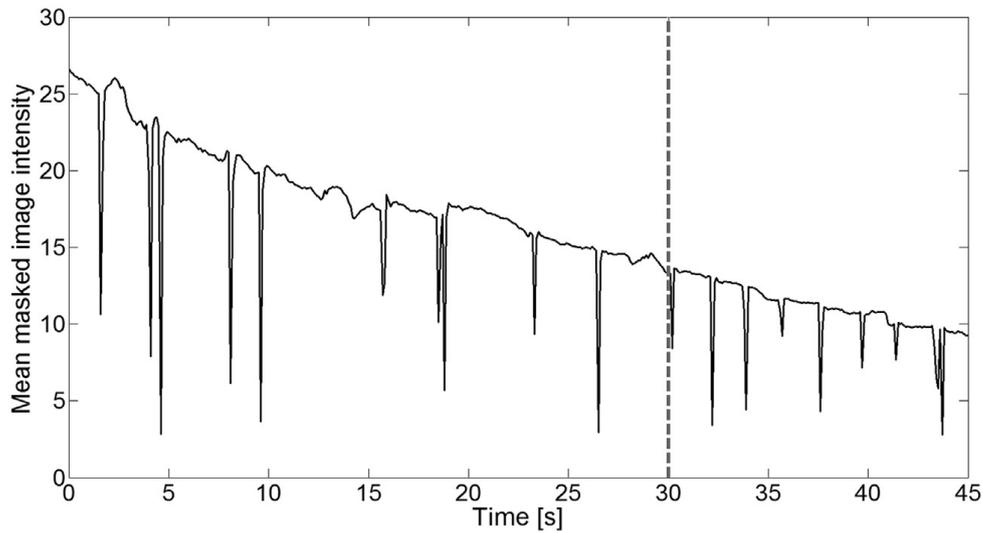

**Fig. 2.** An exemplary curve of fluorescence intensity decay (black line), with 30 s margin marked, acquired for one of the subjects.

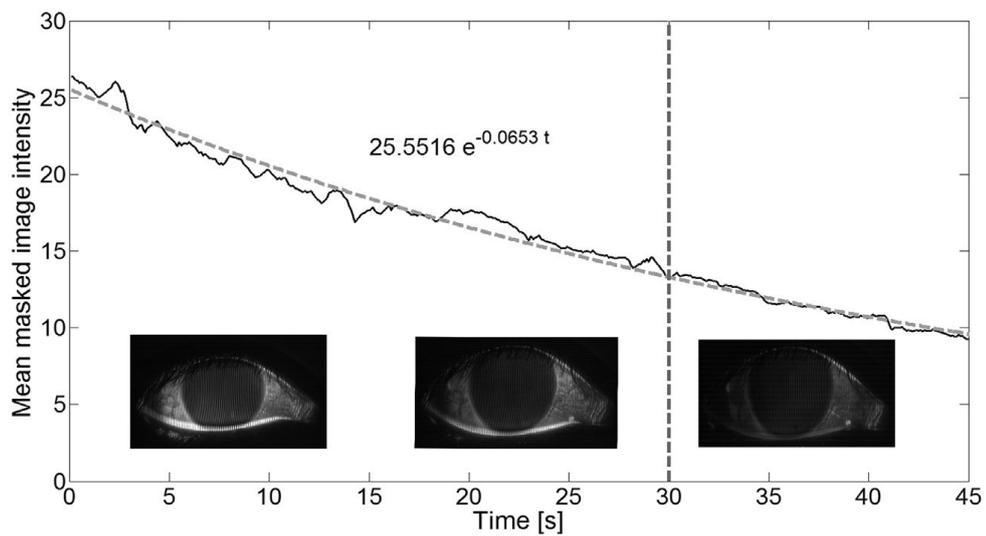

**Fig. 3.** An exemplary curve of fluorescence intensity decay after removal of the signal artefact due to blinks (black line), fitted with exponential curve (grey line) and 30 s margin marked, acquired for one of the subjects. Insets include frames at 4s, 17s, and 31 s of the recording.

The time period for assessing TFWR was set arbitrarily reflecting the pragmatic aspects of the data acquisition. At the same time the exponential decay profiles are available for the entire phase of observed tear film dynamics.

The relatively low mean image intensity (approximately 30–70 AU) may be caused by the fact that the image was acquired from the whole surface of the eye, which contains a highly intense image of the tear meniscus and the image of fringes contrasted with the dark area. In addition, only one from the two illuminating diodes was illuminated during measurements. Using two diodes may have induced reflex tearing caused by the high intensity of the light source (made to be used for very short measurements).

**RESULTS**

All measurements were performed at the same time of the day. The repeatability of the proposed TFWR technique was assessed. Ten randomly selected volunteers from the group of subjects were measured several times. Firstly, TFWR was assessed in the same manner, as described in the Methodology section. Subsequently, the eye was carefully rinsed with saline solution to wash the remaining fluorescent dye and a 10-min break was followed before taking another measurement. Following this, a short measurement of the remaining fluorescein was performed for a period of 10 s. In all cases, no fluorescein was visible in the tear film before taking TFWR measurements. In total, eight measurements were performed per subject. There was a statistically significant positive correlation between the mean value of TFWR and its standard deviation ($r^2 = 0.806$, $p < 0.001$) indicating that the repeatability suffers with greater mean TFWR, which ranged between 1% and 42%. For the smallest TFWR of 1.13% and for the highest TFWR of 40.01% the repeatability was shown to be 27.44% and 24.87%, respectively. Mean repeatability was 28.13 ± 9.59%, with the highest value of 49.01% for a relatively low TFWR, equal to 5.06%.

For all subjects, the mean TFWR was 39% ± 23%, ranging from 1.4% to 83%. The range of these values indicates that TFWR is highly subject-dependent. Fig. 4. shows a histogram of TFWR.

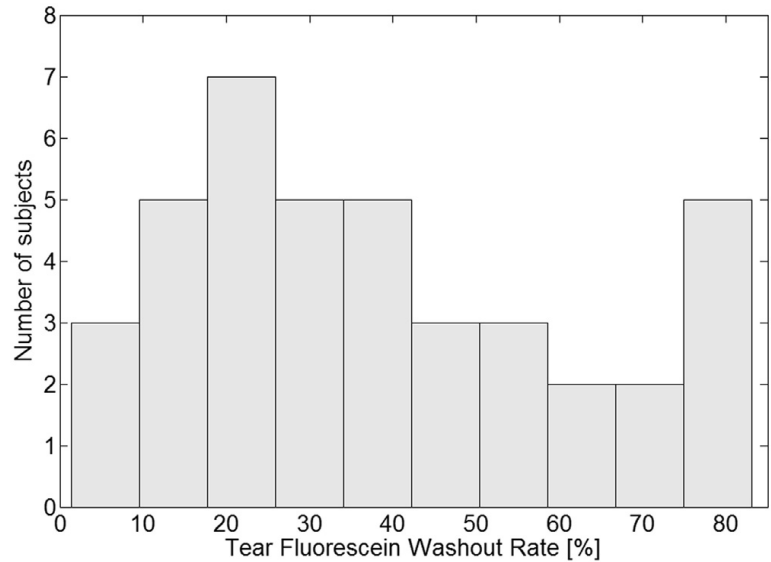

**Fig. 4.** The histogram of tear fluorescein washout rates achieved in the study.

The TFWRs were tested for normality and the null hypothesis was not rejected (Jarque-Bera test, p= 0.291 and Lilliefors test, p = 0.310). Table 1 shows a summary of the data collected (average values, standard deviations, and ranges) and Table 2 shows all the correlation coefficients between TFWR and other measures of tear film including McMonnies questionnaire score, time between subject's awaking and the beginning of the measurements, FTBUT, tear meniscus height and blink frequency.

**Table 1.** Mean, standard deviation and range of the measured parameters.

| Parameter | Mean ± standard deviation | Range |
|---|---|---|
| Age [years] | 31.8 ± 14.2 | [21, 70] |
| McMQ | 7.55 ± 5.21 | [1, 20] |
| Δt [h] | 3.66 ± 1.52 | [1, 6.5] |
| FTBUT [s] | 14.05 ± 7.96 | [5, 30] |
| TMH [mm] | 0.22 ± 0.07 | [0.1, 0.4] |
| Blink frequency [1/min] | 17.10 ± 7.30 | [6, 30] |
| TFWR [%] | 39.17 ± 22.97 | [2, 83] |

McMQ = McMonnies questionnaire score, Δt = time between subject's awakening and the beginning of the measurements, FTBUT = fluorescein tear film break-up time, TMH = tear meniscus height, TFWR = tear fluorescein wash-out rate assessed with fluorescein profilometry

**Table 2.** Pearson correlation coefficients ($r^2$) and statistical significance (p) between Tear Film Washout Rate (TFWR) assessed with fluorescein profilometry and other measures of tear film

| Age | McMQ | Δt | FTBUT | TMH | Blink Freq. |
|---|---|---|---|---|---|
| **$r^2$ = 0.004** | $r^2$ = 0.214 | $r^2$ = 0.009 | $r^2$ = 0.136 | $r^2$ = 0.015 | $r^2$ = 0.006 |
| **p = 0.348** | **p = 0.001** | p = 0.275 | **p = 0.009** | p = 0.225 | p = 0.311 |

McMQ = McMonnies questionnaire score, Δt = time between subject's awakening and the beginning of the measurements, FTBUT = fluorescein tear film break-up time, TMH = tear meniscus height

Statistically significant correlations were found with McMonnies questionnaire score ($r^2$ = 0.214, p = 0.001) and FTBUT ($r^2$ = 0.136, p = 0.009).

It should be noted that the sample size was not estimated before commencing the study as no reasonable information regarding parameter variability was available. However, post-hoc analysis, conducted for 90% power at the 5% significance level, found that the chosen sample of forty subjects could assess differences of about 11% for profilometry based TFWR.

**DISCUSSION**

The main goal was to introduce the results of the first stage of test on a new technique of early phase tear dynamics assessment with fluorescein profilometry and discuss its potential and limitations. This technique is not time-consuming (one measurement lasts up to 1 min), easy to perform, utilises low concentration of fluorescent dye and is performed with a commercially available clinical instrument. The technique can be used to analyse changes occurring in the early phase of tear film dynamics due to high resolution of the images captured and is not limited by the effect of corneal permeability. Use of moistened florets ensures sterility and safety. The pilot study (results not reported) also shows the potential of this method to observe fluid flow in the tear meniscus with blink (Krehbiel flow), however this potential application needs further investigation.

As previously described biphasic characteristics of tear dynamics after dye instillation [1,5,6,10,11] were not observed during this study. It may be assumed that TFWR is more associated with the early-phase of tear clearance rather than the slow, basal phase observed during fluorophotometry.

TFWR was correlated with TFBUT and dry eye symptoms assessed with the questionnaire but not with the other considered measures of tear film dynamics including age, day time, tear meniscus height and blink frequency. The fact that TFWR was weakly correlated with TFBUT does not necessarily mean that the two methods are equivalent and could be substituted. As noted in the introduction, the tear dynamics consists of several factors. TFBUT mostly concerns tear film retention on the surface and evaporation while TFWR relates to tear film distribution, turnover and the elimination through the nasolacrimal system.

It has been established that changes in blink rate lead to significant changes in tear dynamics. Blink rate should be taken into consideration in fluorophotometric measurements [6]. However, our results show that TFWR was not correlated with blink frequency. The study by Wu et al. [24] showed that, in some cases, there were rapid increases in tear turnover rate in the absence of increased tear meniscus height due to rapid blinking. Such effect was not observed in this study. Despite the statistical significance the values of correlation coefficient for TFWR are low. However, it is not an uncommon feature of generally used test of tear film dynamics [25]. After solving the problem of simultaneous blink control and monitoring, choosing the most appropriate amount and concentration of the dye to observe the best image for tear dynamics observation the method may be more repeatable.

A limitation of this study is the lack of control of precise volume and concentration of fluorescein instilled with a strip [26]. In light of this, the methodology would be similar to that used in the assessment of TFBUT for which multiple measurements are required.

It should be noted that the results obtained with fluorescein profilometry should not be directly compared with those obtained from fluorophotometry. The differences between values of TFWR, and those reported in the literature for TTR assessed with fluorophotometry are evident, as well as differences between the amount of time necessary to observe very low image intensity with fluorescein profilometry (less than 1 min) and to follow fluorescein dilution with fluorophotometry (up to 30 min). Possible correlation between measures of tear film dynamics assessed with fluorescein profilometry and fluorophotometry or different types of fluorescein clearance also need further investigation.

Summarising, profilometry based measurement of tear fluorescein wash-out rate could provide a mode for assessing tear film dynamics.

**CONFLICT OF INTEREST**

None

**ACKNOWLEDGEMENT**

This project has received funding from the European Union's Horizon 2020 research and innovation programme under the Marie Skłodowska-Curie grant agreement No 642760.

**REFERENCES**

[1] A. Tomlinson, S. Khanal, Assessment of tear film dynamics: quantification approach, Ocular Surf. 3 (2) (2005) 81–95.

[2] C.G. Begley, R.L. Chalmers, L. Abetz, K. Venkataraman, P. Mertzanis, B.A. Caffery,

C. Snyder, T. Edrington, D. Nelson, T. Simpson, The relationship between habitual patient-reported symptoms and clinical signs among patients with dry eye of varying severity, Investig. Ophthalmol. Visual Sci. 44 (11) (2003) 4753–4761.

[3] K.K. Nichols, J.J. Nichols, G.L. Mitchell, The lack of association between signs and symptoms in patients with dry eye disease, Cornea 23 (8) (2004) 762–770.

[4] C.S. de Paiva, S.C. Pflugfelder, Tear clearance implications for ocular surface health, Exp. Eye Res. 78 (3) (2004) 395–397.

[5] J.D. Nelson, Simultaneous evaluation of tear turnover and corneal epithelial permeability by fluorophotometry in normal subjects and patients with keratoconjunctivitis sicca (KCS), Trans. Am. Ophthalmol. Soc. 93 (1995) 709.

[6] W. Webber, D. Jones, Continuous fluorophotometric method of measuring tear turnover rate in humans and analysis of factors affecting accuracy, Med. Biol. Eng. Comput. 24 (4) (1986) 386–392.

[7] H. Mochizuki, M. Yamada, S. Hatou, K. Tsubota, Turnover rate of tear-film lipid layer determined by fluorophotometry, Br. J. Ophthalmol. 93 (11) (2009) 1535–1538.

[8] A. Macri, S. Pflugfelder, Correlation of the Schirmer 1 and fluorescein clearance tests with the severity of corneal epithelial and eyelid disease, Arch. Ophthalmol. 118 (12) (2000) 1632–1638.

[9] P. Prabhasawat, S.C. Tseng, Frequent association of delayed tear clearance in ocular irritation, Br. J. Ophthalmol. 82 (6) (1998) 666–675.

[10] E. Kuppens, T. Stolwijk, R. de Keizer, J. Van Best, Basal tear turnover and topical timolol in glaucoma patients and healthy controls by fluorophotometry, Investig. Ophthalmol. Visual Sci. 33 (12) (1992) 3442–3448.

[11] J.A. van Best, J.M. del Castillo Benitez, L.-M. Coulangeon, Measurement of basal tear turnover using a standardized protocol, Graefe's Arch. Clin. Exp. Ophthalmol. 233 (1) (1995) 1–7.

[12] K.-P. Xu, K. Tsubota, Correlation of tear clearance rate and fluorophotometric assessment of tear turnover, Br. J. Ophthalmol. 79 (11) (1995) 1042–1045.

[13] W. Webber, D. Jones, P. Wright, Fluorophotometric measurements of tear turnover rate in normal healthy persons: evidence for a circadian rhythm, Eye 1 (5) (1987) 615–620.

[14] A.A. Afonso, L. Sobrin, D.C. Monroy, M. Selzer, B. Lokeshwar, S.C. Pflugfelder, Tear fluid gelatinase B activity correlates with IL-1α concentration and fluorescein clearance in ocular rosacea, Investig. Ophthalmol. Visual Sci. 40 (11) (1999) 2506–2512.

[15] A. Macri, M. Rolando, S. Pflugfelder, A standardized visual scale for evaluation of tear fluorescein clearance, Ophthalmology 107 (7) (2000) 1338–1343.

[16] K. Tsubota, M. Kaido, Y. Yagi, T. Fujihara, S. Shimmura, Diseases associated with ocular surface abnormalities: the importance of reflex tearing, Br. J. Ophthalmol. 83(1) (1999) 89–91.

[17] A.A. Afonso, D. Monroy, M.E. Stern, W.J. Feuer, S.C. Tseng, S.C. Pflugfelder, Correlation of tear fluorescein clearance and Schirmer test scores with ocular irritation symptoms, Ophthalmology 106 (4) (1999) 803–810.

[18] K.-P. Xu, Y. Yagi, I. Toda, K. Tsubota, Tear function index: a new measure of dry eye, Arch. Ophthalmol. 113 (1) (1995) 84–88.

[19] S.C. Pflugfelder, A. Solomon, D. Dursun, D.-Q. Li, Dry Eye and Delayed Tear Clearance: 'a Call to Arms', Lacrimal Gland, Tear Film, and Dry Eye Syndromes 3, Springer, 2002, pp. 739–743.

[20] A. Jordan, J. Baum, Basic tear flow: does it exist? Ophthalmology 87 (9) (1980) 920–930.

[21] L. Sorbara, T. Simpson, S. Vaccari, L. Jones, D. Fonn, Tear turnover rate is reduced in patients with symptomatic dry eye, Cont. Lens Anterior Eye 27 (1) (2004) 15–20. [22] X. Zheng, T. Kamao, M. Yamaguchi, Y. Sakane, T. Goto, Y. Inoue, A. Shiraishi, Y. Ohashi, New method for evaluation of early phase tear clearance by anterior segment optical coherence tomography, Acta Ophthalmol. (Copenh.) 92 (2) (2014) e105–e111.

[23] D.R. Iskander, P. Wachel, P.N. Simpson, A. Consejo, D.A. Jesus, Principles of operation, accuracy and precision of an Eye Surface Profiler, Ophthalmic Physiol. Opt. 36 (3) (2016) 266–278.

[24] Z. Wu, C.G. Begley, N. Port, A. Bradley, R. Braun, E. King-Smith, The effects of increasing ocular surface stimulation on blinking and tear Secretion Ocular stimulation effects on blinking and tear secretion, Investig. Ophthalmol. Visual Sci. 56 (8) (2015) 4211–4220.

[25] B.D. Sullivan, L.A. Crews, B. Sönmez, F. Maria, E. Comert, V. Charoenrook, A.L. de Araujo, J.S. Pepose, M.S. Berg, V.P. Kosheleff, Clinical utility of objective tests for dry eye disease: variability over time and implications for clinical trials and disease management, Cornea 31 (9) (2012) 1000–1008.

[26] A.M. Abdul-Fattah, H.N. Bhargava, D.R. Korb, T. Glonek, V.M. Finnemore, J.V. Greiner, Quantitative in vitro comparison of fluorescein delivery to the eye via impregnated paper strip and volumetric techniques, Optomet. Vision Sci. 79 (7) (2002) 435–438.